\documentclass[manuscript]{aastex}

\usepackage{natbib}
\usepackage{color}
\usepackage{tabularx}

\usepackage{longtable}
\usepackage{multirow}

\slugcomment{International Astronomical Union Proceedings Series}
\shorttitle{GW optical follow up}
\shortauthors{Yang et al.}

\begin{document}
 \title{Gravitational wave optical counterpart searching based on GRAWITA and DLT40 project during LIGO O2 run}

\author{Sheng Yang,$\!$\altaffilmark{1,2} on behalf of GRAWITA and DLT40 project}

\begin{abstract}
The identification of the electromagnetic(EM) counterpart of gravitational wave(GW) trigger in the sky localization is a very difficult task because of the large uncertainty. Two complementary approaches are used in order to search for EM counterpart of GW signal with a typical large sky localization uncertainty: wide-field tilling search on high probability GW region, e.g. Gravitational Wave Inaf Team(GRAWITA) project or pointed search of selected galaxies in high probability GW region, e.g. Distance Less Than 40 Mpc survey(DLT40) project. 

\end{abstract}

 
\altaffiltext{1}{Department of Physics, University of California, 1 Shields Avenue, Davis, CA 95616-5270, USA}
\altaffiltext{2}{INAF Osservatorio Astronomico di Padova, Vicolo dell’Osservatorio 5, I-35122 Padova, Italy}

\section{Introduction}
GW, one of the fundamental predictions of general relativity, were finally detected on 14 September 2015 by the two Laser Interferometer gravitational-wave Observatory (LIGO) interferometers during the first LIGO run(O1)\citep{Abbott16A}. Soon after, the second GW source was detected on 24 November 2015\citep{Abbott16B}. GW is a very important and powerful tool to do astronomy, astrophysics and cosmology while to find EM counterpart of the GW events, which is so-called 'multi-messenger astronomy', will allow a precise localization of the GW source that will help streamline and enhance GW events by reducing the number of free parameters, or at least significantly restrict the range of their values that must be explored. 

In the first, historic detection of GW150914, the two LIGO interferometers could achieve only a poor localization and hence EM counterpart searches had to scan a large area ($10^3$ square deg). With the beginning of joint LIGO and Virgo\citep{Acernese2015} runs currently planned for the Fall 2016\citep{ligo1}, the simultaneous operations of 3 interferometers will secure a substantial reduction of the error areas. Nevertheless the error areas will still be dozens to hundred of square degrees and only cameras with very large field of view(FoV) can cover them entirely with a limited number of tiled pointings. One may wonder if it's possible to search the GW region with small telescopes whose answer is yes. \citeauthor{galaxy_cat_1, galaxy_cat_2, galaxy_cat_3} shows that searching the LIGO region based on galaxy catalog could greatly increases the chance of imaging an EM counterpart and the issue is to have a galaxy catalogue with a good completeness. GLADE\footnote{{\bf http://aquarius.elte.hu/glade/}} catalog, which is considered as the most complete galaxy catalogue so far, can have a good completeness up to 70 Mpc, which means that we can use pointed strategy within this distance while once GW trigger happened further, the only choice would be the large FoV 'blind' search.

During the second run(O2) of the LIGO and Virgo Interferometer, a GW signal consistent with a binary neutron star coalescence(BNS) was detected on 2017 August 17th (GW170817)\citep{Abbott17A}, quickly followed by a coincident short gamma-ray burst (GRB170817a\citep{fermigcn}) trigger by the Fermi satellite\citep{fermi}. About ten hours later, the DLT40 supernova search reported the discovery of a new optical transient, DLT17ck, as one of the six optical follow-up groups which independently detected this coincident optical kilonovae, 2017 gfo/sss17a\citep{Abbott17B,sss17a}. DLT17ck was indeed cooling down and getting dimmer much faster than any other supernova(SN) we ever observed after monitoring the source for several days. Soon after, the spectra of this fast transient taken by some follow-up groups, including GRAWITA, proved that it's an r-process kilonovae which is characterized by rapidly expanding ejecta with spectral features similar to those predicted by current models. The era of multi-messenger astronomy has truly begun. 

\section{DLT40-like pointed galaxy search}
Once the GW trigger occur in the local universe where the galaxy catalogue is still available, the galaxy pointing follow-up strategy, like DLT40, would be feasible. 

\begin{figure*}
\begin{center}
\includegraphics[height=0.3\textwidth, width=0.45\textwidth]{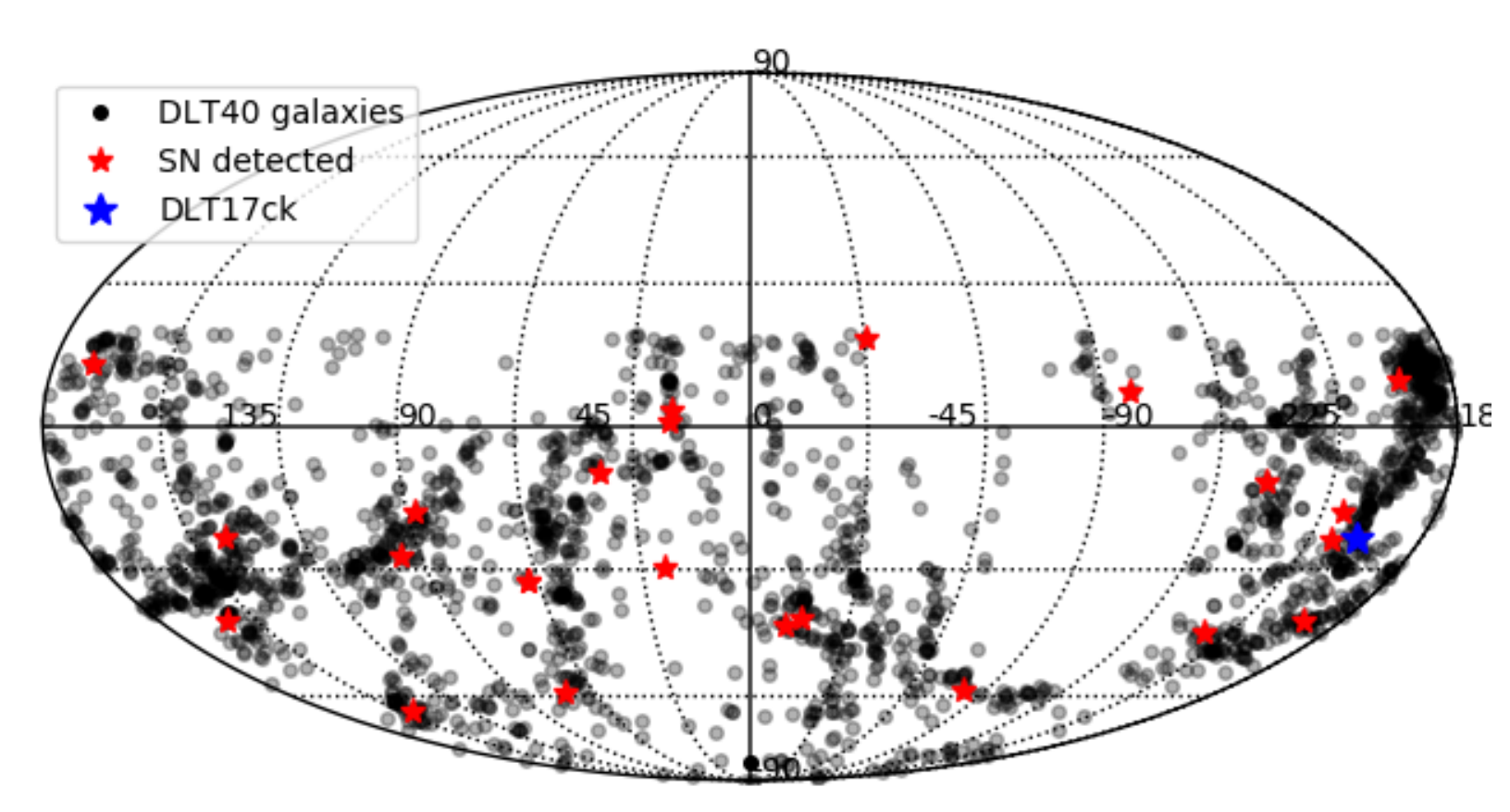}
\includegraphics[height=0.55\textwidth, width=0.45\textwidth]{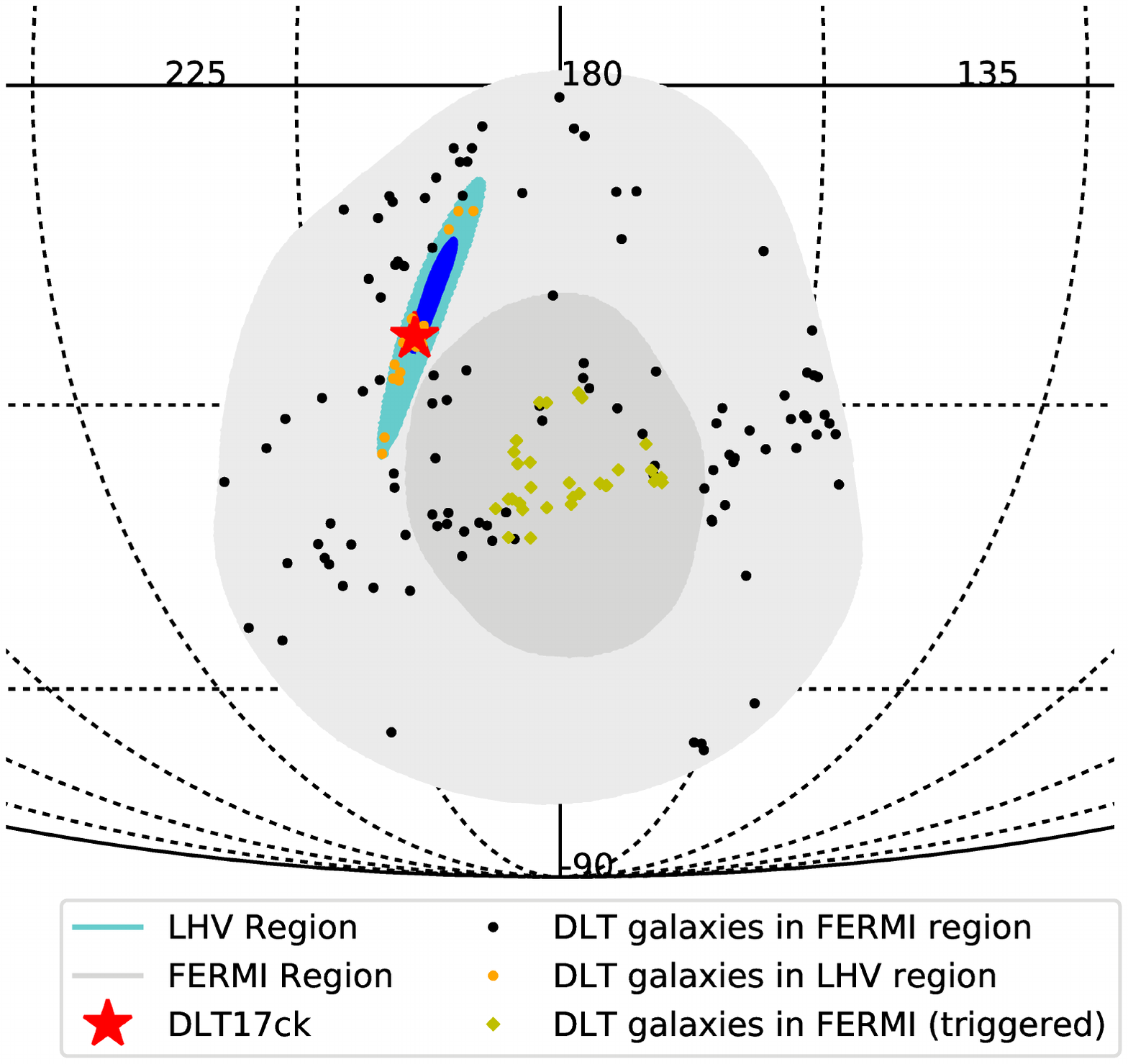}
\caption{{\bf Left panel}: DLT40 galaxy catalogue map with all SN detected by DLT40 so far and DLT17ck\citep{yang17}. {\bf Right panel}: The sky map region of GW170817 over-imposed on the Fermi GBM trigger GRB108017A. The red star marks the location of DLT17ck and the host galaxy NGC 4993\citep{Valenti17}.}
\label{fig:dlt40}
\end{center}
\end{figure*}

DLT40 is a one day cadence supernova search in nearby galaxies aim to discover SN within 24 hours from explosion since this is the time when we can learn the most on the physics of the explosion. Within hundreds of square degrees, indeed, there is a limited number of nearby galaxies\citep{Gehrels}. The quick response of the Prompt telescopes (that we use for DLT40) is ideal to quickly observe the LIGO/Virgo region. A second advantage is that, since DLT40 is daily monitoring nearby galaxies\footnote{DLT40 is monitoring a sample of 2000 galaxies that have been selected among the galaxies of the Gravitational Wave Galaxy Catalogue (GWGC, \cite{GWGC}). Starting from the GWGC catalogue, we selected only galaxies with recessional velocity $V\le 3000 km*s^{-1}$ (corresponding to $D\le40$ Mpc), with a declination ($Dec\le+20$ deg), absolute magnitude ($MB\le-18$ mag), and Milky Way extinction ($A_{V}\le0.5$ mag). Figure \ref{fig:dlt40} Left panel shows all the 2000 DLT40 galaxy samples together with all SN found so far and the kilonoava, DLT17ck.}, we have useful information on those galaxies before detecting any variability. As soon as a new GW event is detected and the sky localization map is released. DLT40 would cross match the localization map with DLT40 galaxy catalog giving highest rank to bright galaxies coincident with high-probability map region. We then observed these galaxies for a period of 2 weeks for burst events and 3 weeks for merge events, which is related to the expected time evolution of optical counterpart of different LIGO events\footnote{burst events may be related to SN explosion that usually peak in two weeks. The merging of two neutron stars should produce an optical counterpart that has been predicted to be visible for few weeks (at least in red bands), but giving the high uncertainty of these models we select a conservative window of 3 weeks.}.

\begin{figure*}
\begin{center}
\includegraphics[height=0.33\textwidth, width=0.45\textwidth]{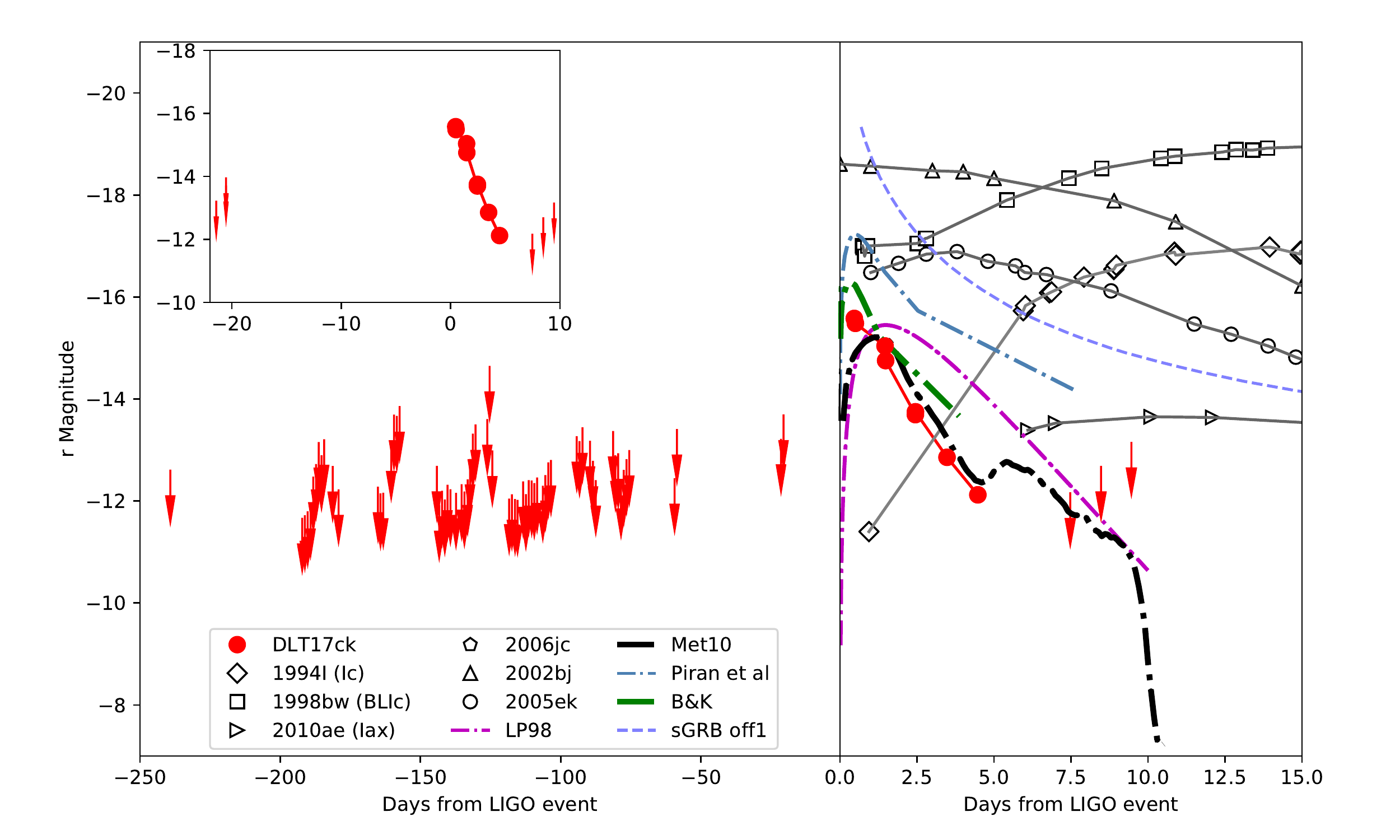}
\includegraphics[height=0.35\textwidth, width=0.45\textwidth]{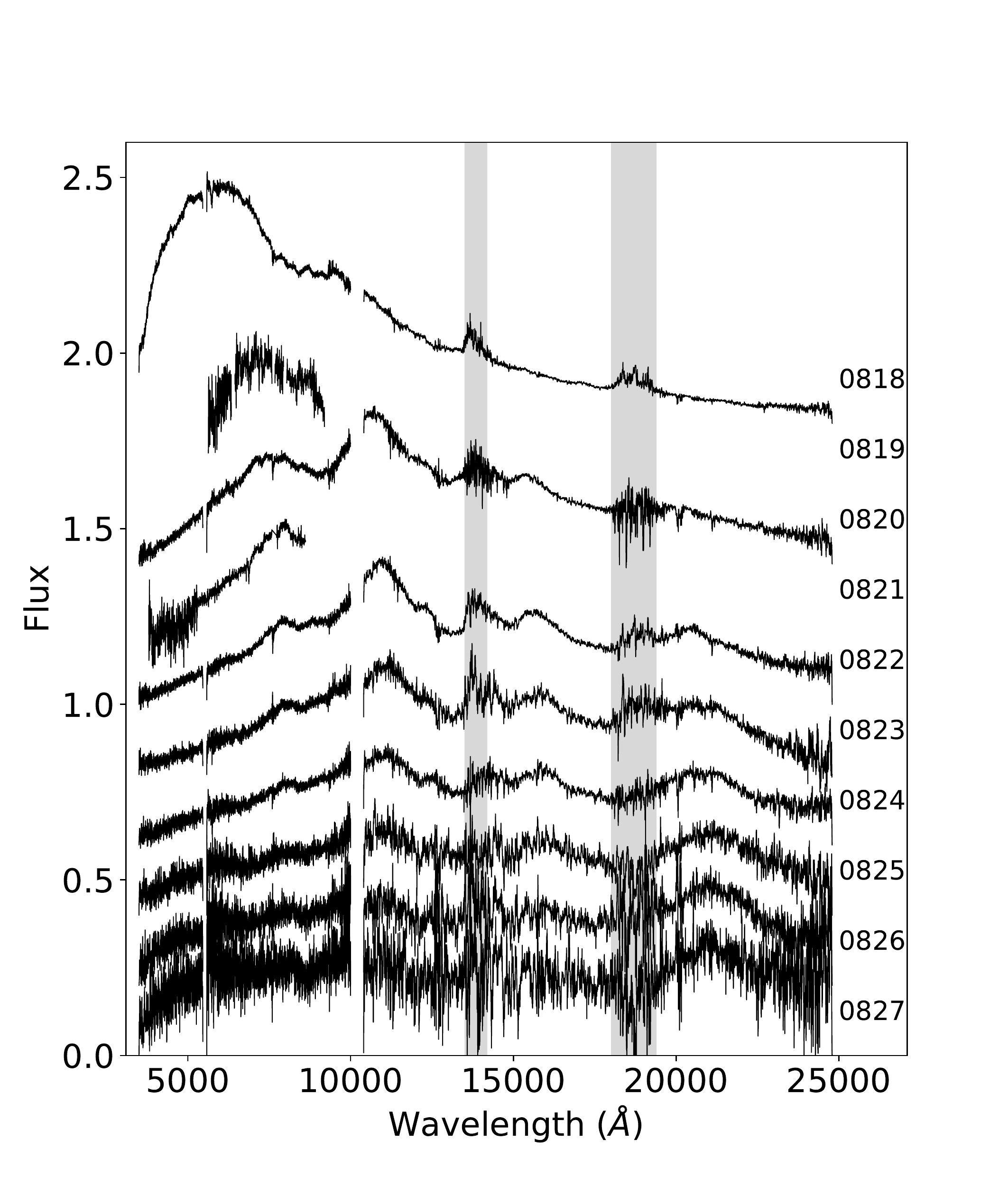}
\caption{{\bf Left panel}: DLT17ck light curve(in red) over plotted with normal or fast-evolving SNe (in gray). Several NS-NS
merger models, scaled to a distance of 40 Mpc, are shown as comparison\citep{Valenti17}. {\bf Right panel}: The sequence of X-shooter, FORS2, and GMOS spectra for SSS17a/DLT17ck\citep{grawita2}.}
\label{fig:lc}
\end{center}
\end{figure*}

\begin{figure*}
\begin{center}
\begin{tabular}{p{0.5\textwidth} p{0.5\textwidth}}
  \vspace{0pt} \includegraphics[height=0.35\textwidth, width=0.45\textwidth]{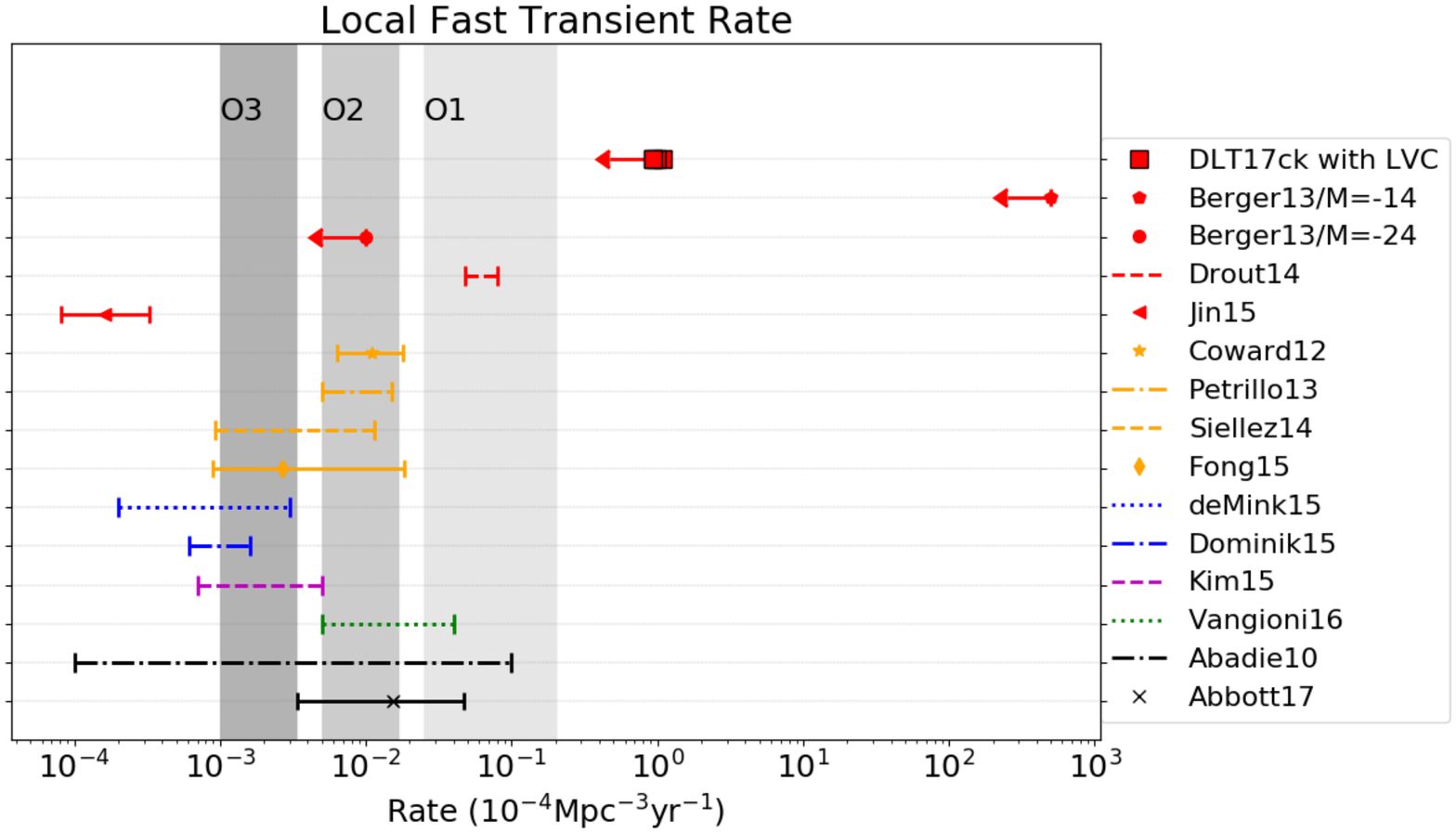} &
  \vspace{20pt} \includegraphics[height=0.25\textwidth, width=0.45\textwidth]{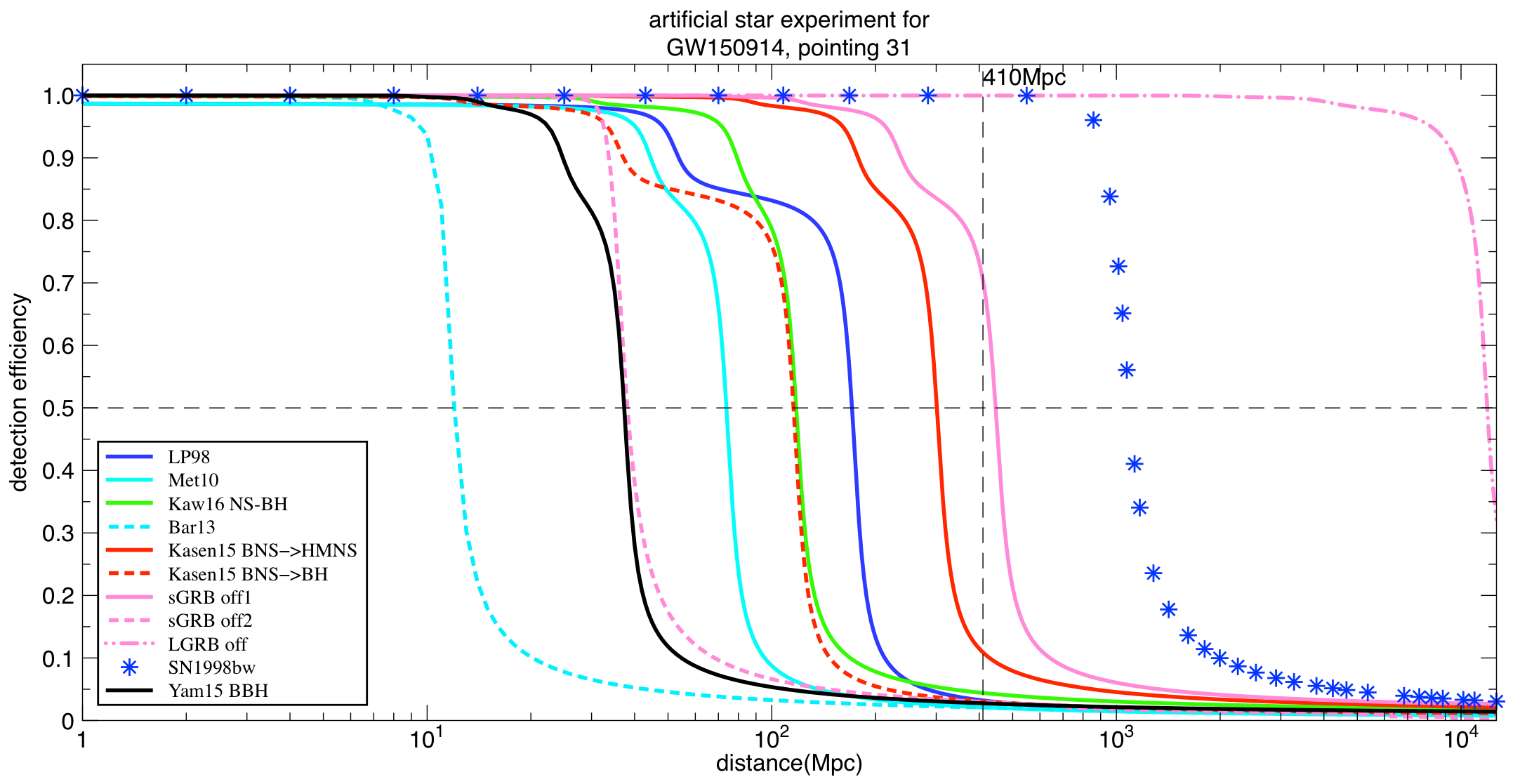}
\end{tabular}
\caption{{\bf Left panel}: DLT40 kilonova rate estimation, compared with the rate of sGRB(orange), BNS merger from stellar evolution(blue), cosmic nucleosyntesis(green), galactic pulsar population(magenta), gravitational waves(black) and fast optical transients(red)\citep{yang17}. {\bf Right panel}: How far VST can reach? With artificial star experiment, we test different kilonova models to estimate the limiting distance of VST images\citep{grawita1}.}
\label{fig:rate}
\end{center}
\end{figure*}

During the LIGO O2 period, on 2017 August 17 23:49:55 UT (11.09 hours after the LVC event GW170817), DLT40 detected DLT17ck, at RA=13:09:48.09 and DEC=-23:22:53.4.6, 5.37W, 8.60S arcsec offset from the center of NGC~4993 \citep[][see Figure \ref{fig:dlt40} Right Panel]{yanggcn}, as one of the six groups who indepandently detected this kilonova\citep{Valenti17}. \cite{yang17} used the observed light curve of DLT17ck(See Figure \ref{fig:lc} Left panel) to constrain the rate of BNS mergers to less than $0.50\,_{-0.04}^{+0.05}\rm{SNuB} / 0.99\,_{-0.15}^{+0.19}\,10^{-4}\,\rm{Mpc^{-3}}\,\rm{yr^{-1}}$ with SNe Ia extinction(See Figure \ref{fig:rate} Left panel). After considering the completeness of the tracing mass and consuming time difference, we conclude that DLT40 need to be oprerated for $\sim18.4 \rm{years}$ in order to independently discover a kilonova without GW information. 

\section{GRAWITA-like large FoV blind search}
Once the GW trigger occur further where the galaxy catalogue completeness is poor, one may consider if the galaxy pointing follow-up strategy is still working and if not, the solution could be the large FoV blind follow-up strategy, like GRAWITA.

\begin{figure*}
\begin{center}
\includegraphics[height=0.25\textwidth, width=0.45\textwidth]{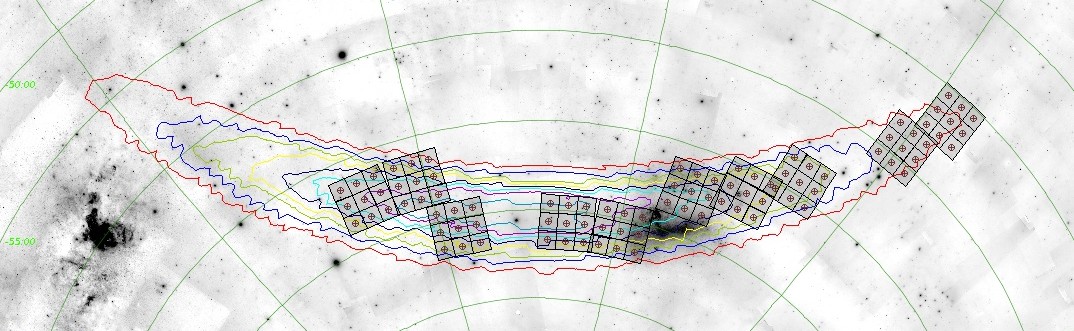}
\includegraphics[height=0.3\textwidth, width=0.45\textwidth]{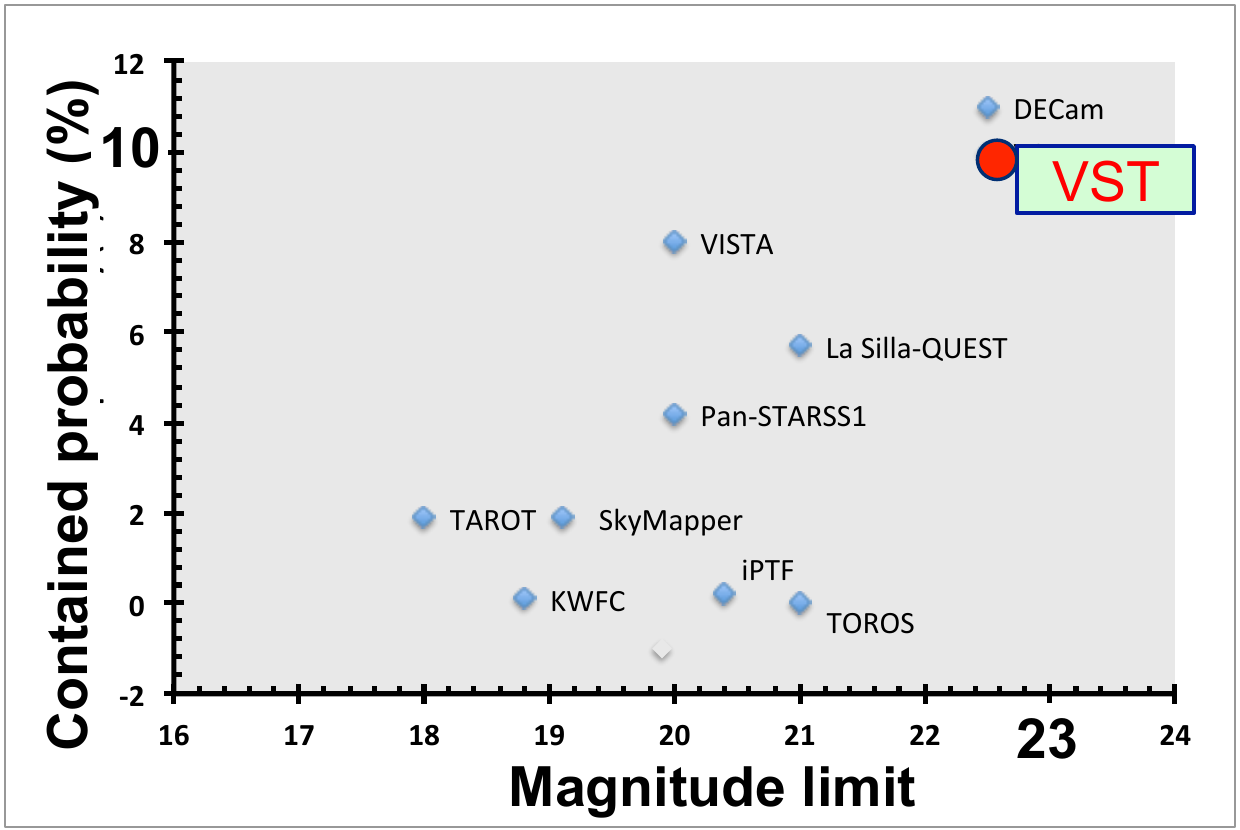}
\caption{{\bf Left panel}: VST 1 square degree survey boxes(90 pointings, 6 epochs) scheduled for GW150914.{\bf Right panel}: VST performance: VST could include 10\% probability of GW map with 22 mag in r band on average\citep{grawita1}.}
\label{fig:grawita}
\end{center}
\end{figure*}

GRAWITA is an EM follow-up team by carrying out follow-up observational campaigns in the optical/NIR bands of the GW trigger observable with the VST, VLT, LBT, TNG, REM ground-based facilities. The VST 2.6m telescope(that GRAWITA used to follow the GW region) with 1 square degree FoV, can survey up to hundreds of square degrees every night and can reach to a limiting magnitude of r=22 mag on average by dithering 2 images with 45s exposure, which is very well designed for blind search following the GW area since it can reach deep enough and cover very large field(see Figure \ref{fig:grawita} Right panel). Also, most of the sources would be visible for GRAWITA at 410Mpc, which is the best value of GW150914 prediction, based on artificial star experiments towards different EM emission models(see Figure \ref{fig:rate} Right panel). Figure \ref{fig:grawita} Left panel is an example by the VST 1 square degree survey boxes scheduled for GW150914. We performed the survey in r band with six epochs while the temporal sampling is chosen allowing for the rapid evolution of the expected GW's counterpart. Our immediate objective is to identify as early as possible all transients in the skymap area and select the best candidates for GW's counterparts. Once an interesting transient detected that can be link to a GW trigger, GRAWITA team and international collaborators would observe this source with an almost daily cadence in spectroscopic way, e.g. GRAWITA contributes to the spectroscopic classification of the kilonova, DLT17ck\citep{grawita2}.

\section{Implications and Future Prospects}
Both strategies have played a very important role so far, in particular for the GW170817 event, no wonder that it's the success for the small FoV follow-up strategy but how about the future? \cite{yang17} has such discussion that: During O3 run, LIGO and VIRGO will be able to detect BNS coalescences out to $65-85$ $\,\rm{Mpc}$\footnote{We consider $85\,\rm{Mpc}$ in the discussion}. Considering all kilonovae detected later would be as bright as DLT17ck, while with the current DLT40 observing strategy, the limiting distance for DLT17ck is $70\,\rm{Mpc}$.  In order to cover the full LIGO/Virgo volume(in case one kilonova was detected in the limiting area), we would need to go $\sim0.4\,\rm{mag}$ deeper, hence increasing the exposure time by a factor of 2.2, which means DLT40 could observe only $\sim230$ galaxies during a single night instead of current $\sim 500$ galaxies. Even if we neglect that the GLADE catalog is only $\sim 85-90\%$ complete in the integrated luminosity up to $85\,\rm{Mpc}$, the number of randomly selecting galaxies within $85\,\rm{Mpc}$ from the GLADE catalog in typical aLIGO/aVirgo regions (30 sq degrees) is $\sim230$, which is more or less the limiting number for DLT40. Therefore, within $85\,\rm{Mpc}$, small telescopes can still play a useful role \footnote{unless DLT17ck turns out to be a particularly bright kilonova}, but the incompleteness of the available catalogs, especially for faint galaxies, may suggest that a wide FoV tilling strategy may be preferred to avoid possible biases in sampling of the stellar population. 

The wide FoV strategy would be the better and sometimes the only choice if the BNS source happened further more. From the GRAWITA wide FoV searching experiences, we found that the difficulty of this strategy would be the large amount of consuming time for running pipelines and choose the best source as soon as possible during thousands of candidates. The parallel computing would increase the computing speed and instead of eyeballing candidates one by one, the machine learning algorithm would be the better choice and sometimes the only choice to rank the candidate list. 

With more interferometers joining in the future(e.g. KAGRA and LIGO-India), the GW sky localization would be more certain and the EM follow-up would become more effective, we believe that more multi-messenger sources would be detected and based on them would allow us to enrich our knowledge of the astrophysics in the multi-messenger astronomy era, thus, the past and the future in the precise cosmology erae.

\bibliographystyle{apj}

\clearpage

\end{document}